\documentclass[twocolumn,pra,aps,superscriptaddress,showpacs]{revtex4-1}
\usepackage{array}
\usepackage{amsmath}
\usepackage{graphicx}
\usepackage{amsmath}
\usepackage{latexsym}
\usepackage{amsfonts}
\usepackage{amssymb}
\usepackage{array}
\usepackage{epsfig}
 \usepackage{bbm}
 \usepackage{color}
 \usepackage{hyperref}

\newcommand{\be}{\begin{equation}}
\newcommand{\ee}{\end{equation}}
\newcommand{\bae}{\begin{eqnarray}} \newcommand{\eae}{\end{eqnarray}}

\def\Tr{{\rm tr}}

\begin{document}

\title{Optimal estimation of joint parameters in phase space}

\author{M. G. Genoni}
\affiliation{QOLS, Blackett Laboratory, Imperial College London, SW7 2AZ, United Kingdom}
\author{M. G. A. Paris}
\affiliation{Dipartimento di Fisica, Universit\`a degli Studi di Milano, I-20133, Milano, Italy}
\author{G. Adesso}
\affiliation{School of Mathematical Sciences, University of Nottingham, Nottingham NG7 2RD, United Kingdom}
\author{H. Nha}
\affiliation{Department of Physics, Texas A \& M University at Qatar, Doha, Qatar}
\author{P. L. Knight}
\affiliation{QOLS, Blackett Laboratory, Imperial College London, SW7 2AZ, United Kingdom}
\author{M. S. Kim}
\affiliation{QOLS, Blackett Laboratory, Imperial College London, SW7 2AZ, United Kingdom}

\date{\today}

\begin{abstract}
We address the joint estimation of the two defining parameters of a
displacement operation in phase space.
In a measurement scheme based on a Gaussian
probe field and two homodyne detectors, it is shown that both conjugated
parameters can be measured below the standard quantum
limit when the probe field is entangled.
We derive the most informative Cram\'er-Rao bound, providing the
theoretical benchmark on the estimation and observe that our scheme is nearly optimal
for a wide parameter range characterizing the probe field.
We discuss the role of the entanglement as well as the relation
between our measurement strategy and the generalized
uncertainty relations.
\end{abstract}

\pacs{42.50.Xa, 03.65.Ta}

\maketitle

\section{Introduction}\label{s:1intro}
One of the most promising avenues for quantum
technology is the use of quantum resources to improve
the sensitivity in the estimation of (not directly observable)
relevant physical parameters, {\em e.g.} for applications in metrology
and sensing \cite{metrology, matteoLQE}.
While the estimation of a single parameter has been extensively
studied from both theoretical and experimental viewpoints \cite{metrology,ExpKonrad}, the
{\it joint} estimation of multiple parameters has not received enough attention 
(notable contributions are \cite{yuen, HelKen, belavkin,holevo,Naga,
Fuji1, Chiri06,You09, Wata10, MonIllu,fabrizio,Cro12}). 
Here we take a further step forward in this direction by providing optimal quantum-enhanced
strategies to estimate the two conjugated parameters characterizing
a paradigmatic and ubiquitous quantum operation, phase-space displacement.
Quantum states of single-mode bosonic CV systems
can be described by quasiprobability distributions on a two-dimensional
real phase space \cite{BarRad}.  The operation of displacing a
state by a phase space vector $(q_0,p_0)$ is represented by
the Weyl displacement operator
\begin{equation}
\hat{D}(q_0,p_0)=\exp(ip_0\hat{q}-iq_0\hat{p})\,,
 \end{equation}
 where $\hat{q}$ and $\hat{p}$ are the two quadrature
 operators  satisfying the canonical commutation relation
 $[\hat{q},\hat{p}]=i\mathbbm{1}$.  A relevant question is, starting
 with a reference state $\varrho_0$ that undergoes an unknown displacement, how
 accurately can we jointly estimate the two conjugate parameters $q_0$ and $p_0$
 of the displacement operator with a measurement on the displaced
 state $\varrho=\hat D (q_0,p_0)\varrho_0\hat{D}^\dag (q_0,p_0)$?  One
 possibility is certainly to use coherent states as initial probe states, followed by
 heterodyne detection as a measurement strategy at the output of the
 displacement transformation \cite{yuen}. This has in fact been the standard
 technique to estimate and measure displacement, as it naturally
 complies with the generalized Heisenberg uncertainty relation
 \cite{Arthurs}.  On the other hand, one may ask whether
entanglement, in the form of Einstein-Podolsky-Rosen (EPR)
correlations \cite{ReidReview},
could lead to a better estimation precision for this task, as it
 was suggested in \cite{matteo2}, as it happens in the
 case of phase estimation in quantum metrology \cite{metrology} 
and for the joint estimation of the 
 temperature and coupling constant in a bosonic dissipative channel \cite{fabrizio}.
 \par
 In this paper we propose and analyze a measurement scheme
 where CV entanglement is used to improve the estimation
precision for this particular and relevant problem. For this purpose,
we limit the analysis to Gaussian states and operations \cite{Gauss}. 
An introduction on LQE for multiple parameters will be given in Sec. \ref{s:2LQE},
while in Sec. \ref{s:3CRB} we will derive the bounds for the 
displacement estimation by considering single- and two-mode Gaussian
states. In Sec. \ref{s:4scheme} we show a simple measurement scheme
involving two-mode squeezed thermal states which achieves the ultimate
bound for different values of squeezing and thermal photons and beats
the standard quantum limit for this kind of estimation. We also
discuss how the performances and the bounds change when one has 
a-priori information about the parameters to be estimated, and when 
the displacement operation presentes an inner uncertainty. In Sec. \ref{s:5ent}
we discuss the role of entanglement, showing how it is always necessary
to beat the classical optimal strategy, and also necessary for symmetric 
probe states. In Sec. \ref{s:6uncrel} we discuss the relationship between
this multi-parameter estimation and the generalized uncertainty relations, 
and finally we end the paper with some concluding remarks.
\section{Multi-parameter local estimation theory} \label{s:2LQE}
How can we know that a given measurement scheme, able to estimate 
certain parameters, is optimal? Is it possible to estimate these parameters
with a better precision? 
In order to answer this question, we make use of tools derived from
local quantum estimation (LQE) theory
\cite{cramerrao,helstrom,BC,matteoLQE}. The purpose of LQE theory is indeed
to determine the ultimate precision achievable and the corresponding
optimal measurement for the estimation of parameters characterizing a
physical quantum system. In particular it has been applied for the
estimation of different quantities, including the quantum phase of a
harmonic oscillator \cite{monrasPhase,PhDiffCV},  CV Gaussian unitary
parameters \cite{GaibaKerrSq}, the amount of quantum correlations of
bipartite quantum states \cite{EntEst,EntEstExp}, and the
coupling constants of different kinds of interactions
\cite{Sar,Hotta,MonrasTeo,Adessoetal,Aspachsetal,Fuji2,Ji,Boixo,fabrizio,MonIllu,escher}.
While most studies so far have been limited to single parameter
estimation, the case of multiple parameters is more complex as
different bounds can be derived for the same setting.  Moreover, as we
will point out later, these bounds are not always achievable, in
particular when one deals with conjugate variables for which a
Heisenberg-type uncertainty relation applies.\\
Let us start by considering the general case, that is a family of
quantum states $\varrho_{\bf z}$ which depend on a set of
$d$ different parameters ${\bf z} = \{ z_\mu \}$,
$\mu = 1, \cdots, d$. One can define the so called
Symmetric Logarithmic Derivative (SLD) and Right Logarithmic
Derivative (RLD) operators for each one of the parameters
involved, respectively as
\begin{align}
\frac{\partial\varrho}{\partial z_\mu} &= \frac{L_\mu^{(S)} \varrho + \varrho L_\mu^{(S)}}{2} \qquad &{\rm SLD}.  \label{eq:SLD} \\
\frac{\partial\varrho}{\partial z_\mu} &=\varrho  L_\mu^{(R)}  \qquad &{\rm RLD}. \label{eq:RLD}
\end{align}
Then one can define the two matrices
\begin{align}
{\bf H}_{\mu\nu} &=  \Tr \left[ \varrho_{\bf z}\frac{ L^{(S)}_\mu L^{(S)}_\nu + L^{(S)}_\nu L^{(S)}_\mu}{2} \right ] \:,  \\
{\bf J}_{\mu\nu} &= \Tr[ \varrho_{\bf z} L_\nu^{(R)} L_\mu^{(R)\dag}].
\end{align}
By defining the covariance matrix elements $V({\bf z})_{\mu\nu}
= E[z_\mu z_\nu ] - E[ z_\mu] E[ z_\nu]$ and
a weight (positive definite) matrix ${\bf G}$, two different 
Cram\'er-Rao bounds holds
\begin{align}
{\rm tr} [ {\bf G V} ]   &\geq \frac1M {\rm tr}[ {\bf G} ({\bf H})^{-1}] \:, \\
{\rm tr} [ {\bf G V} ]   &\geq  \frac{{\rm tr}[{\bf G}{\rm Re}({\bf J}^{-1})] 
+  {\rm tr}[|{\bf G}{\rm Im}({\bf J}^{-1})|]  }M \:, 
\end{align}
where ${\rm tr}[A]$ is the trace operation on a finite dimensional matrix
$A$ and $M$ is the number of measurements performed. 
We observe that if we choose ${\bf G}=\mathbbm{1}$, we obtain
the two bounds on the sum of the variances of the parameters involved
\begin{align}
\sum_\mu {\rm Var} (z_\mu) &\geq \frac{{\sf B}_{\rm S}}M := \frac1M {\rm tr}[{\bf H}^{-1}]  \:,\\
\sum_\mu {\rm Var} (z_\mu) &\geq \frac{{\sf B}_{\rm R}}M := \frac{{\rm tr}[{\rm Re}({\bf J}^{-1})] +
{\rm tr}[|{\rm Im}({\bf J}^{-1})|]  }M \:.
\end{align}
The matrices
${\bf H}$ and ${\bf J}$ are called respectively the Symmetric Logarithmic Derivative (SLD)
\cite{helstrom} and Right Logarithmic Derivative (RLD) \cite{yuen,belavkin,holevo,Naga,Fuji1}
quantum Fisher information matrices. 
Neither the SLD bound ${\sf B}_{\rm S}$, nor the RLD bound ${\sf B}_{\rm R}$
on the sum of the variances
are in general achievable \cite{HelKen}. The first one could not be achievable because it
corresponds to the bound obtained by measuring optimally and simultaneously
each single parameter and this is not possible when the optimal
measurements do not commute. At the same time the RLD bound could not be achievable
because the optimal estimator does not always correspond to a proper quantum
measurement (that is a proper positive operator valued measure).
Moreover which one of these bounds is more informative, that
is which one is higher and then tighter, depends strongly on the
estimation problem considered. One can then define the {\em most informative}
Cram\'er-Rao bound, 
$$
{\sf B}_{\rm MI} = \max \{   {\sf B}_{\rm S}, {\sf B}_{\rm R}  \} \:,
$$
obtaining the single inequality
$$
\sum_\mu {\rm Var} (z_\mu) \geq \frac{{\sf B}_{\rm MI}}{M}.
$$
\subsection{Cram\'er-Rao bounds with a-priori information}
Similar bounds can be obtained in the case where one has a certain a-priori information
regarding the distribution
on the parameters one wants to estimate. Let us assume that the a-priori information is 
described by a probability distribution
$\mathcal{P}_{{\sf prior}} ({\bf z})$. One can define a Fisher-information
matrix of the a-priori distribution as
\begin{align}
{\bf A}_{\mu\nu} = \int d{\bf z} \: \mathcal{P}_{{\sf prior}} ({\bf z}) \left( 
\frac{\partial \log \mathcal{P}_{{\sf prior}} ({\bf z})}{\partial z_\mu}\right)
\left(\frac{\partial \log \mathcal{P}_{{\sf prior}} ({\bf z})}{\partial z_\nu}\right).
\end{align}
The new Cram\'er-Rao bounds that take into account this a-priori information, 
will read
\begin{align}
{\rm tr} [ {\bf G V} ]   &\geq  \frac1M {\rm tr}[ {\bf G} ({\bf H}+{\bf A})^{-1}] \:, \\
{\rm tr} [ {\bf G V} ]   &\geq \frac1M \left( {\rm tr}[{\bf G}{\rm Re}(({\bf J}+{\bf A})^{-1})]
+  {\rm tr}[|{\bf G}{\rm Im}(({\bf J}+{\bf A})^{-1})|]   \right)
\end{align}
and, for ${\bf G}=\mathbbm{1}$,
\begin{align}
\sum_\mu {\rm Var} (z_\mu) &\geq \frac{{\sf B}_{\rm S}({\bf \Delta})}{M} 
:= \frac1M{\rm tr}[({\bf H}+{\bf A})^{-1}]  \:, 
\label{eq:SLDprior} \\
\sum_\mu {\rm Var} (z_\mu)  &\geq \frac{{\sf B}_{\rm R}(\boldsymbol{\Delta})}{M} 
 := \frac1M\left( {\rm tr}[{\rm Re}(({\bf J}+{\bf A})^{-1})] \: + \right.\nonumber \\
& \left. \qquad \qquad + {\rm tr}[|{\rm Im}(({\bf J}+{\bf A})^{-1})|] \right) \:. \label{eq:RLDprior}
\end{align}
Here ${\bf \Delta}$ denotes a vector of parameters characterizing the
prior information at our disposal.
\subsection{Evaluation of the RLD Fisher information}
In the following we give some details about the derivation
of the RLD Fisher information when the RLD operator cannot
be evaluated directly.
Let us suppose that the derivative with respect to every parameter
has the following form
\begin{align}
\frac{\partial \varrho}{\partial z_\mu} = \varrho L_\mu^{(a)} + B_\mu \varrho. \label{eq:deriv}
\end{align}
To obtain the RLD operator $L_\mu= L_\mu^{(a)} + L_\mu^{(b)}$
as defined in Eq. (\ref{eq:RLD}) we have
 to find the operator $L_\mu^{(b)}$ such that (assuming that $\varrho^{-1}$ exists)
 \begin{align}
B_\mu \varrho &= \varrho L_\mu^{(b)} \:. \\
 L_\mu^{(b)} &= \varrho^{-1} B_\mu \varrho \:.
 \end{align}
Then, after some algebra, we can express the elements of the
RLD Fisher information matrix as
\begin{align}
J_{\mu\nu} &= \Tr[\varrho L_\nu L_\mu^{\dag}] \nonumber \:,  \\
&=  \Tr[\varrho L_\nu^{(a)} (L_\mu^{(a)})^{\dag} ] + \Tr[\varrho B_\mu^{\dag} L_\nu^{(a)}]+
\nonumber \\
& \:\: + \Tr[\varrho (L_\mu^{(a)})^{\dag} B_\nu] + \Tr[B_\nu \varrho^2 B_\mu^{\dag} \varrho^{-1}] \:.
\label{eq:Jmunu}
\end{align}
\section{Cram\'er-Rao bounds for displacement estimation} \label{s:3CRB}
Let us now consider a generic probe state (pure or mixed) $\varrho_0$, which is displaced
by the operator $\hat{D}(q_0,p_0)$ to the
state $\varrho = \hat{D}(q_0,p_0) \varrho_0  \hat{D}^{\dag}(q_0,p_0)$.
In the following we derive explicit formulas for the SLD and RLD
Fisher information matrices.
Let us start by considering the SLD Fisher information for a given 
probe state whose diagonal form reads $\varrho_0 = \sum_n p_n |\phi_n\rangle\langle\phi_n|$.
One proves that in our case the SLD operator in Eq. (\ref{eq:SLD})
satisfies the property
$L_\mu^{(S)} = \hat{D}(q_0,p_0) \mathcal{L}_\mu \hat{D}^{\dag}(q_0,p_0)$, where
\begin{align}
\mathcal{L}_\mu = 2i \sum_{n\neq m} \langle G_\mu \rangle_{nm} \frac{p_n - p_m}{p_n + p_m}
|\phi_n\rangle\langle \phi_m|
\end{align}
with  $\mu, \nu=\{q_0, p_0\}$,
$ \langle G_\mu \rangle_{nm} = \langle \phi_n | G_\mu | \phi_m\rangle$
and where $G_{q_0}=\hat{p}, G_{p_0}=-\hat{q}$
are the generators of the two orthogonal displacements.
Then the SLD Fisher information matrix elements read
\begin{align}
H_{\mu\nu} &= \frac{1}{2} \Tr [ \varrho_0 ( \mathcal{L}_\mu \mathcal{L}_\nu +
\mathcal{L}_\nu \mathcal{L}_\mu)] \:,  \\
&= 2 \sum_{s\neq t} p_s \left( \frac{p_s - p_t}{p_s + p_t} \right)^2 \left( \langle G_\mu \rangle_{st}
\langle G_\nu\rangle_{ts} + \langle G_\nu \rangle_{st} \langle G_\mu\rangle_{ts} \right) \:.
\end{align}
Let us consider now more in detail the case of the RLD Fisher information.
By differentiating $\varrho$ with respect to the parameters $\mu = \{q_0, p_0\}$, we
obtain formulas resembling  Eq. (\ref{eq:deriv}), where
$L_\mu^{(a)} = B_\mu^{\dag} = - i G_\mu$. Then, starting
from Eq. (\ref{eq:Jmunu}), and by observing
that
\begin{align}
\hat{D}^\dag(\mu) \hat{p} \hat{D}(\mu) &= \hat{p} - p_0 \\
\hat{D}^\dag(\mu) \hat{q} \hat{D}(\mu) &= \hat{q} + q_0,
\end{align}
we can express, after some algebra, 
the elements of the Fisher information matrix in terms of the
generators of the displacement as
\begin{align}
J_{\mu\nu} &= \Tr[G_\nu \varrho_0^2 G_\mu \varrho_0^{-1}] + \Tr[\varrho_0 G_\nu G_\mu] 
- 2 \Tr[\varrho_0 G_\mu G_\nu]\,.\notag
\end{align}
We notice that the Fisher matrices
do not depend on the values of the parameters to be estimated
and that the only elements that are involved are the probe state
and the generators of the two transformations.
\subsection{Most informative bounds for single- and two-mode probe states}
The most general single-mode Gaussian state with zero initial
displacement can be written as $\varrho_0 = S(r) \nu_N S(r)^{\dag}$
where $\nu_N = \frac{1}{N+1}\sum_n \frac{N}{N+1}|n\rangle\langle n|$ 
is a thermal state and $S(r)=\exp\{-\frac{r}{2}(a^{\dag\:2}-a^2)\}$ is the single-mode
squeezing operator. Notice that every single-mode squeezed state
evolving in a noisy dissipative channel can always be written in this
form, which makes this treatment important for actual implementations \cite{Seraparis}.
In this case the two bounds ${\sf B}_{\rm S}$ and ${\sf B}_{\rm R}$ 
are evaluated and
the most-informative for the single-mode case 
${\sf B}_{\rm MI}^{(1)}$ is found to be equal to the 
RLD bound, yielding
\be
{\sf B}_{\rm MI}^{(1)}(r,N)  = (2N+1)\cosh{2r}+1\:.
\ee
For zero squeezing the results obtained by Yuen and Lax is recovered \cite{yuen}.
Moreover one can verify that this bound is achieved for any value 
of squeezing and thermal photons by performing an heterodyne measurement.
In general we observe that the bound grows with $N$ and $r$. 
It is thus clear that single-mode squeezing is not useful for displacement
estimation and the optimal measurement setup involving single-mode Gaussian
probe states and heterodyne detection corresponds to using the vacuum
(or any coherent state) as a probe field. The corresponding
bound is denoted by 
\begin{align}
{\sf B}_{\rm sql} ={\sf B}_{\rm MI}^{(1)}(0,0) = 2 \:, \label{eq:SQL}
\end{align}
as to the standard quantum limit (SQL). We note that the
SQL does not depend at all on the mean energy
of the probe coherent state: by increasing the mean photon number of the coherent states one does not obtain any enhancement in the estimation precision.\\
Let us focus now on the more interesting two-mode case, where
the displacement operator is applied only on one of the two-modes.
The probe state corresponds to a two-mode squeezed thermal state, which is
an archetype of the (possibly noisy) Gaussian entangled states:
\begin{align}
\varrho_0 =
 \hat{S}_2(r) (\nu_N\otimes \nu_N) \hat{S}_2^\dag(r), \label{eq:TMS}
\end{align}
where $\hat{S}_2(r)=\exp\{ r(\hat{a}^\dag \hat{b}^\dag - \hat{a}\hat{b})\}$ is
the two-mode squeezing operator.
The two bounds can also be straightforwardly evaluated, obtaining
\begin{align}
{\sf B}_{\rm S}^{(2)}(r,N) &= \frac{2 N +1}{ \cosh{2r}} \label{eq:twoSLDB}\\
{\sf B}_{\rm R}^{(2)}(r,N) &= \frac{4 N (1+N)}{(2 N+1)\cosh{2r} -1} \label{eq:twoRLDB} .
\end{align}
 Both are increasing functions of the average number
 of thermal photons $N$ and decreasing functions of
 the squeezing parameter $r$ (and thus of the entanglement
 of the probe state). In this case which bound is the most informative
 depends on the actual values of $r$ and $N$.
 Comparing Eqs (\ref{eq:twoSLDB}) and (\ref{eq:twoRLDB}), when
 $\cosh(2r) < 2N+1$, ${\sf B}_{\rm S} < {\sf B}_{\rm R}$.  Thus we
 define a threshold value for the squeezing  as $
 r_{\rm ths}= \frac{1}{2} \cosh^{-1}(2N + 1)$,
and the most informative bound  reads
\be
{\sf B}_{\rm MI}^{(2)}(r,N) =
\left\{ \begin{array}{c}
{\sf B}_{\rm R}^{(2)}(r,N) \qquad \textrm{for}\: r<r_{\rm ths} \\
{\sf B}_{\rm S}^{(2)}(r,N) \qquad \textrm{for}\: r\geq r_{\rm ths}.
\end{array}
\right.
\ee
We notice that for $N=0$ the most informative bound coincides with the
SLD bound, while if we increase the value of $N$ and for small values of the squeezing
parameter $r$, the most informative bound turns out to be the RLD bound.
By inspecting the most informative bound ${\sf B}_{\rm MI}^{(2)}$,
we notice that for different values of the parameters the bound is smaller than
the SQL bound ${\sf B}_{\rm sql}$. One may then wonder if by using 
entangled probe states one can achieve a better result; in the next
section we indeed present a simple measurement scheme, outperforming 
the classical single-mode strategy. 
\begin{figure}[tb]
\includegraphics[width=0.99\columnwidth]{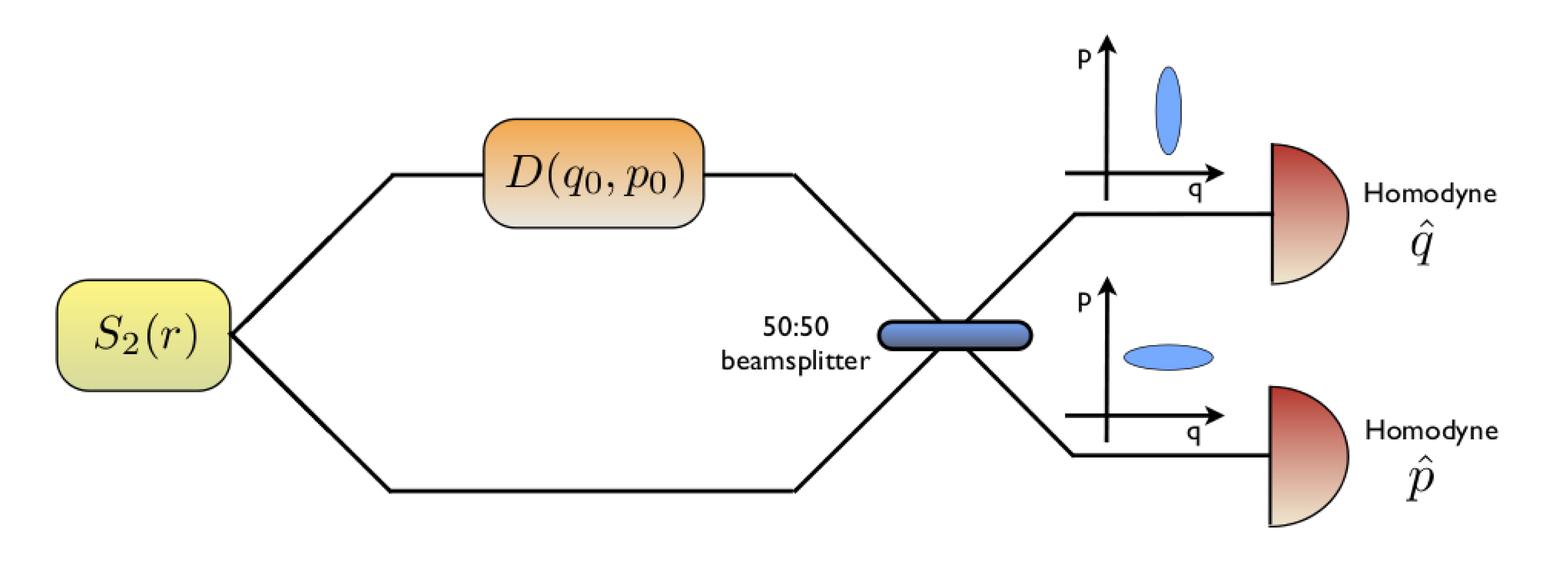}
\caption{
\label{f:schemeA}
(Color on-line) Measurement scheme for the estimation of the displacement
given a two-mode squeezed probe state. After the displacement operation,
the modes are mixed in a balanced beam splitter and then
orthogonal homodyne measurements are performed on the output modes.
}
\end{figure}
\section{Near-optimal measurement scheme} \label{s:4scheme}
As pointed out in the previous section, if we consider coherent states
as probe states and then, after the displacement operation, we perform
a heterodyne measurement, we achieve the SQL ${\sf B}_{\rm sql}$.
On the other hand, the bounds obtained for entangled probe states
suggest that the SQL can be in principle overcome. 
Indeed we now illustrate the two-mode measurement scheme able to beat
bound ${\sf B}_{\rm sql}$ and to achieve the optimality for different values of the 
parameters characterizing the probe state. The scheme is pictured in Fig. \ref{f:schemeA}.
It clearly resembles the CV version 
of the dense coding protocol \cite{DCBraunstein} and was already
suggested for the estimation of displacement \cite{matteo2}.
The probe state corresponds to a two-mode squeezed thermal state (\ref{eq:TMS}).
The displacement operator is applied on one mode after which
the two modes are mixed at a balanced beam splitter. Then the output fields of the beam splitter are described by the density operator
\begin{align}
\varrho^\prime &= \hat{U}_{\rm bs} \varrho
\hat{U}_{\rm bs}^\dag = \varrho_1^\prime \otimes \varrho_2^\prime
\end{align}
where $\hat{U}_{\rm bs}=\exp\{{\pi\over 4}(\hat{a}\hat{b}^\dag-\hat{a}^\dag\hat{b})\}$ is the beam splitter operator \cite{BarRad}, $\varrho=\hat{D}(q_0,p_0)\varrho_0 \hat{D}^\dag(q_0,p_0)$ is
the state after the displacement, and
\begin{align}
\varrho_1^\prime &=\hat{D}(q_0^\prime,p_0^\prime)\hat{S}(r)\nu_N \hat{S}^\dag(r) \hat{D}^\dag(q_0^ \prime,p_0 \prime) \\
\varrho_2^\prime &=\hat{D}(q_0^\prime,p_0^\prime)\hat{S}(-r)\nu_N \hat{S}^\dag(-r) \hat{D}^\dag(q_0^\prime,p_0^\prime)
\end{align}
with $q_0^\prime = \frac{q_0}{\sqrt{2}}$, $p_0^\prime = \frac{p_0}{\sqrt{2}}$ \cite{msKbcS}.
The output state
is a tensor product of two states squeezed in orthogonal directions and both
displaced by the rescaled values $q_0^\prime$ and $p_0^\prime$.  One performs a homodyne measurement of the quadrature $\hat{p}$ on the state
$\varrho_1^\prime$ and of the quadrature $\hat{q}$ on $\varrho_2^\prime$, obtaining
respectively the parameter values $q_0$ and $p_0$. As the states are squeezed
in orthogonal directions, the two variances approach exponentially to zero by increasing
the squeezing parameter $r$ as ${\rm Var}(q_0) = {\rm Var}(p_0) = \left( 2N + 1 \right)e^{-2r}$:
the higher is the squeezing the more precise is the estimation.
The sum of the two variances is
\be
{\rm Var}(q_0) + {\rm Var}(p_0) = 2 \left( 2N+1\right)e^{-2r} \geq {\sf B}_{\rm MI}^{(2)}(r,N)
\label{eq:twomode}
\ee
One can observe that we obtain for the two-parameter estimation the
same optimal scaling in terms of the degree of squeezing, as the
one obtained for the single-parameter displacement estimation in
\cite{Munro}; in particular for a pure two-mode squeezed
state ($N=0$), one achieves for large squeezing the Heisenberg limit 
scaling $1/\bar{N}$, where $\bar{N}=\sinh^2 r$ denotes the mean number of photons.
Comparing Eq.~(\ref{eq:twomode}) with Eq.~(\ref{eq:SQL}), 
it is clear that this scheme can outperform the
single-mode strategy.  For $N=0$, as long as squeezing is non-zero, we can estimate the parameters better than the SQL suggests. For
$N\neq 0$, if the field exhibits two-mode squeezing, that is, if
it is squeezed stronger than the following threshold
\be
r > r_{\rm sql} (N) = \frac14 \ln\left( 1+4N + 4 N^2\right) \label{eq:thr}
\ee
we can beat the SQL. 

If we now compare the obtained results to the most-informative bounds derived 
in the previous section, we observe that the sum of the variances 
${\sf E}(r,N)= {\rm Var}(q_0) + {\rm Var}(p_0)$ in Eq. (\ref{eq:twomode}) 
is, as expected, bounded from below by ${\sf B}_{\rm MI}^{(2)}(r,N)$. However
one may wonder if, in some range of parameters, the scheme becomes
(nearly) optimal, that is if the bound is almost saturated.
\begin{figure}[tb]
\includegraphics[width=0.7\columnwidth]{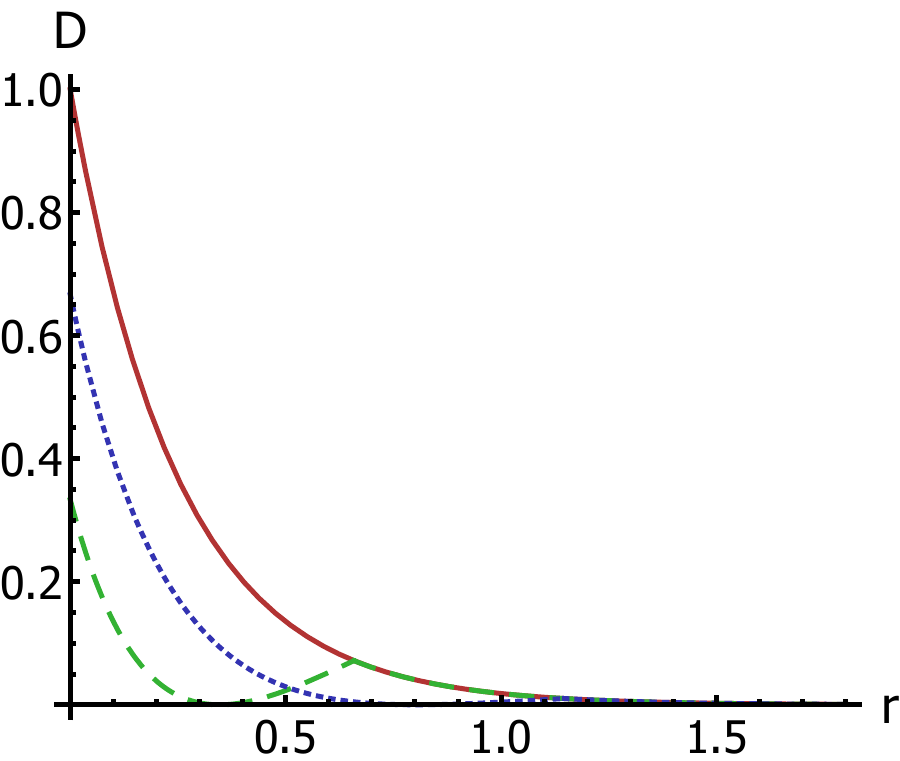}
\caption{
\label{f:DH}
(color on-line) Renormalized difference ${\sf D}(r,N)$ between the sum of the variances for the estimation of displacement with the double-homodyne scheme and the most informative bound
${\sf B}_{\rm MI}^{(2)}(r,N)$,
as a function of the squeezing parameter $r$ and for different values of thermal
photons: continuous-red line  $N=0$;  dashed-green line $N=0.5$; dotted-blue line $N=2$.
}
\end{figure}
In Fig. \ref{f:DH}
we plot the quantity
$$
{\sf D}(r,N) = \frac{{\sf E}(r,N) - {\sf B}_{\rm MI}^{(2)}(r,N)}{{\sf B}_{\rm MI}^{(2)}(r,N)}
$$ 
for different values of $N$ and as a function of the squeezing $r$.
One observes that by increasing the squeezing parameter $r$, our
scheme is optimal with ${\sf D}(r,N)\simeq 0$. For lower values of $r$,
we notice that ${\sf D}(r,N)$ is not always monotonically decreasing;
this is because the most informative bound
changes between the RLD bound ${\sf B}_{\rm R}^{(2)}(r,N)$ and
the SLD bound ${\sf B}_{\rm S}^{(2)}(r,N)$, as explained before.
As remarked before, the measurement scheme we present resembles the
CV version of the dense coding protocol \cite{DCBraunstein}; however, while 
the dense coding protocol requires more than $4\: \textrm{dB}$ of squeezing to outperform
single-mode strategies, our estimation strategy outperforms the SQL for
{\it any} value of two-mode squeezing at the input. 
\subsection{Estimation with a-priori information}
Let us consider now the case where we have some prior distribution
on the parameters we want to estimate. In particular, for the sake of 
simplicity we consider the parameters taken randomly with the following
a-priori probability distribution:
$$
\mathcal{P}_{{\sf prior}}(q_0,p_0) = \mathcal{G}_{0, \Delta}(q_0) \, \mathcal{G}_{0,\Delta}(p_0),
$$
where $\mathcal{G}_{\mu,\sigma^2}(x)$ denotes a Gaussian distribution, centered 
at $\mu$ and with
variance $\sigma^2$. The objective is to minimize the average precision
one gets on the estimation of these random parameters;
we can then evaluate
the bounds in Eqs. (\ref{eq:SLDprior}) and (\ref{eq:RLDprior}), for different input states 
and strategies. By considering coherent states as the input, the most-informative RLD 
bound will be equal to
\begin{align}
{\sf B}_{{\sf SQL}}(\Delta) := \frac{2 \Delta^2}{1+ \Delta^2} . \label{eq:prioriSQL}
\end{align}
If we rather consider a two-mode squeezed thermal state $\varrho_0 = S_2(r) \nu_N \otimes \nu_N
S_2^\dagger(r)$ as the probe, we obtain the following bounds:
\begin{align}
{\sf B}_{\rm S}^{(2)}(r,N,\Delta) &= \frac{2 (2 N +1) \Delta^2}{2N+1+2 \Delta^2 
\cosh{2r}} \label{eq:twoSLDB}\\
{\sf B}_{\rm R}^{(2)}(r,N,\Delta) &= \frac{4 N (1+N) \Delta^2}{2 N(1+N)+\Delta^2[(2 N+1)\cosh{2r} -1]} \label{eq:twoRLDB} .
\end{align}
All these bounds decrease with the decreasing $\Delta$. 
We note that all the 
bounds discussed before can be re-obtained
by taking the limit of flat a-priori distribution ($\Delta \rightarrow \infty$).
If one fixes the value of $\Delta$, 
one can define the most-informative bound ${\sf B}_{\rm MI}^{(2)} 
= \max \{   {\sf B}_{\rm S}^{(2)}, {\sf B}_{\rm R}^{(2)}  \}$, and compare it
with the corresponding SQL bound. 
One could then ask which is the optimal measurement strategy, and if the 
precision obtained saturates the most informative bounds for different possible
probe states. Let us start by considering input coherent states: as proved in 
\cite{yuen}, the optimal cheating strategy corresponds to multiply the heterodyne 
outcomes by a factor
\be
K_c = \frac{\Delta^2}{1+\Delta^2}. \label{eq:Kc}
\ee
It is indeed easy to check that with this choice the obtained
averaged variances are equal to the SQL limit ${\sf B}_{\sf SQL}(\Delta)$
derived in Eq. (\ref{eq:prioriSQL}) (notice that this is also the optimal 
choice used in \cite{braunTLP,HamTLP} to derive the classical benchmark for teleportation 
of coherent states).
Let us consider the general case where, 
for given values of $q_0$ and $p_0$, the variances
obtained with a certain measurement strategy are equal and 
do not depend on the parameters themselves, that is
$${\rm Var}(q_0) = {\rm Var}(p_0) := {\rm Var}_0 \:.
$$ 
One can prove that
the scaling factor that minimizes the average sum of the variances is
equal to
\be
K_{\rm min} = \frac{ \Delta^2}{{\rm Var}_0 + \Delta^2},
\ee
and the obtained result is
$$
\langle\:  {\rm Var}_K(q_0) + {\rm Var}_K(p_0) \: \rangle = \frac{ 2 {\rm Var}_0 \: \Delta^2}{{\rm Var}_0 + \Delta^2} \:,
$$
where $\langle \: \cdot \: \rangle$ denotes the average on the
a-priori distribution.
This is also the case for the two-mode squeezed thermal states 
considered before. Of course the scaling factor in this case
depends on the probe state parameters, since
\be
{\rm Var}_0 = (2 N +1)e^{-2r}.
\ee
If this information is not available, one can always adopt the
coherent states optimal strategy and use the scaling factor
$K_c$ in Eq. (\ref{eq:Kc}), which does not depend on the
input state, obtaining
\be
\langle\:  {\rm Var}_{K}(q_0) + {\rm Var}_K(p_0) \: \rangle = \frac{ \Delta^2 (1 + \Delta^2 {\rm Var_0})}{(1+\Delta^2)^2}.
\ee
\begin{figure}[h!]
\includegraphics[width=0.45\columnwidth]{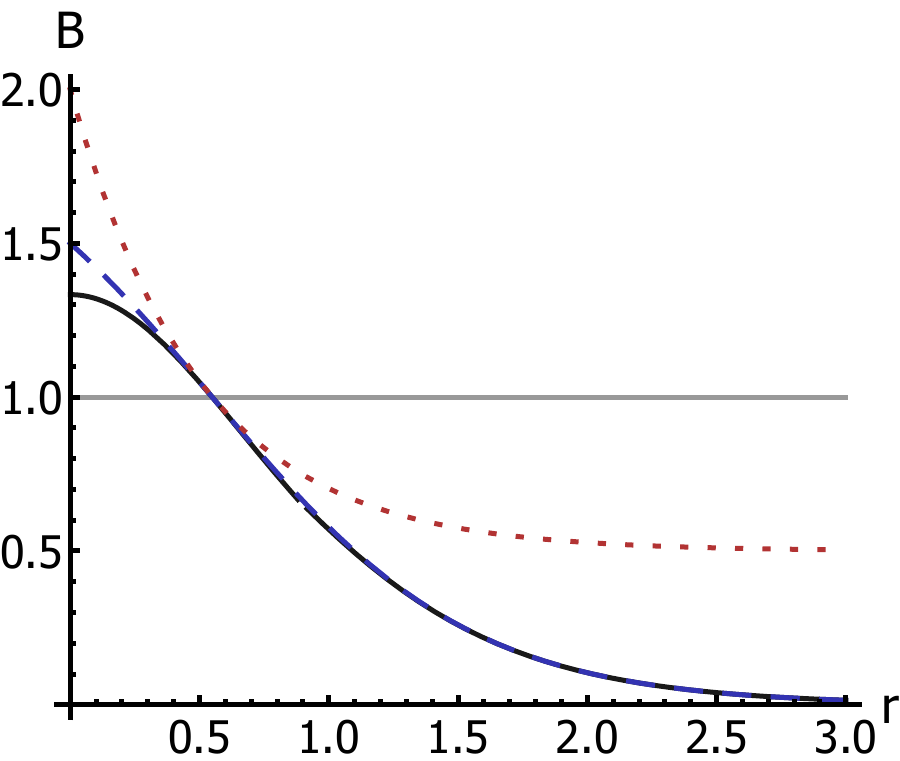}
\includegraphics[width=0.45\columnwidth]{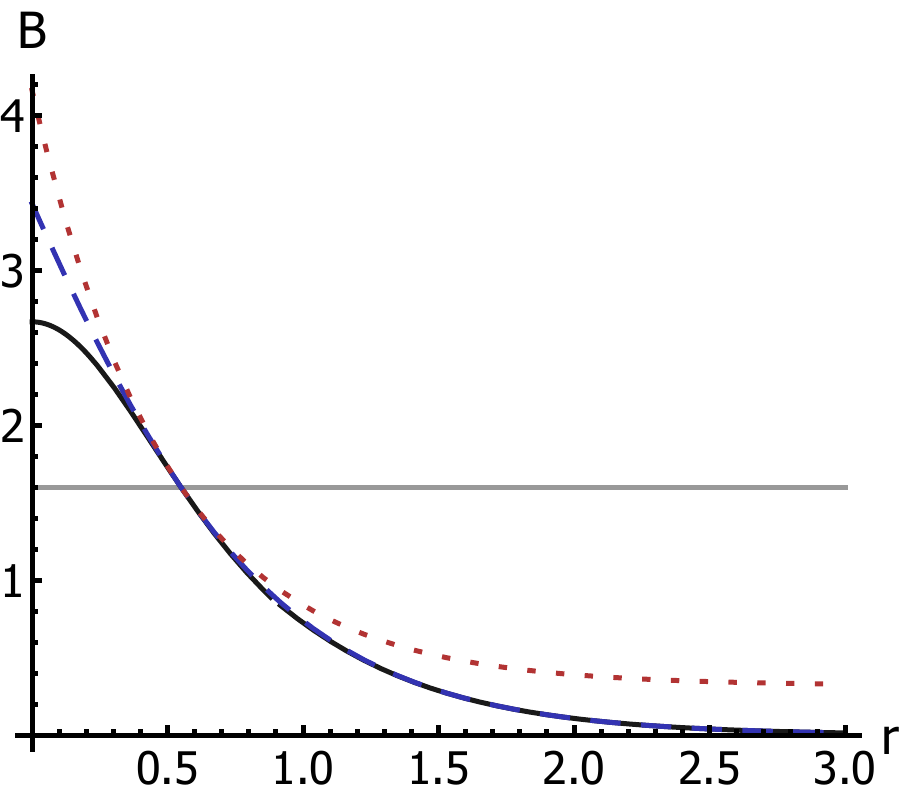} \\
\includegraphics[width=0.45\columnwidth]{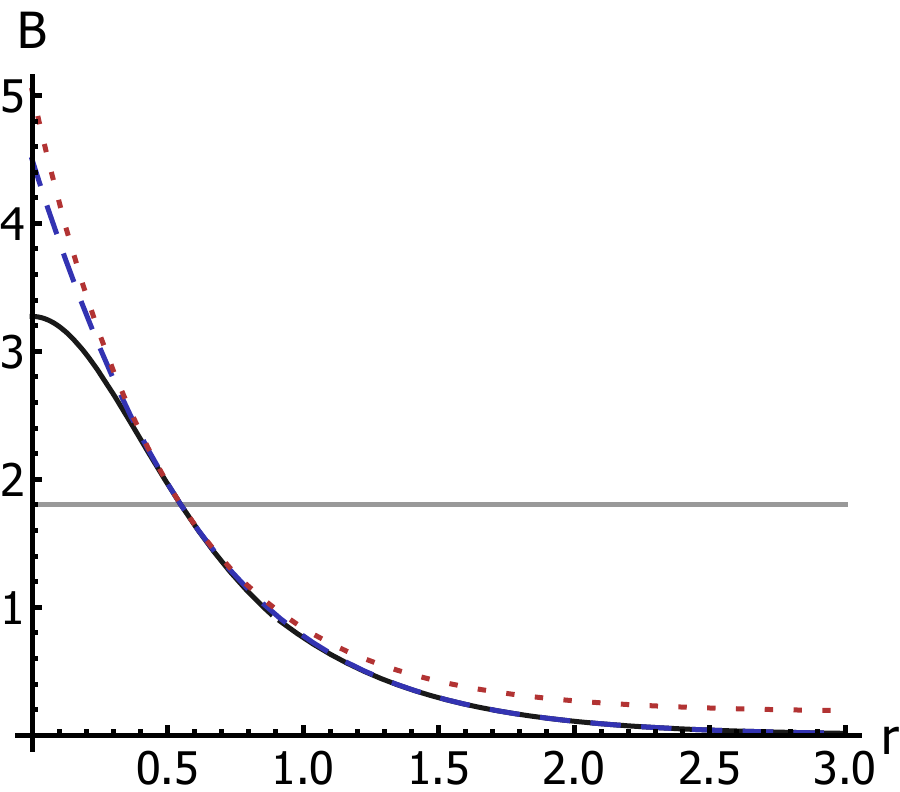}
\includegraphics[width=0.45\columnwidth]{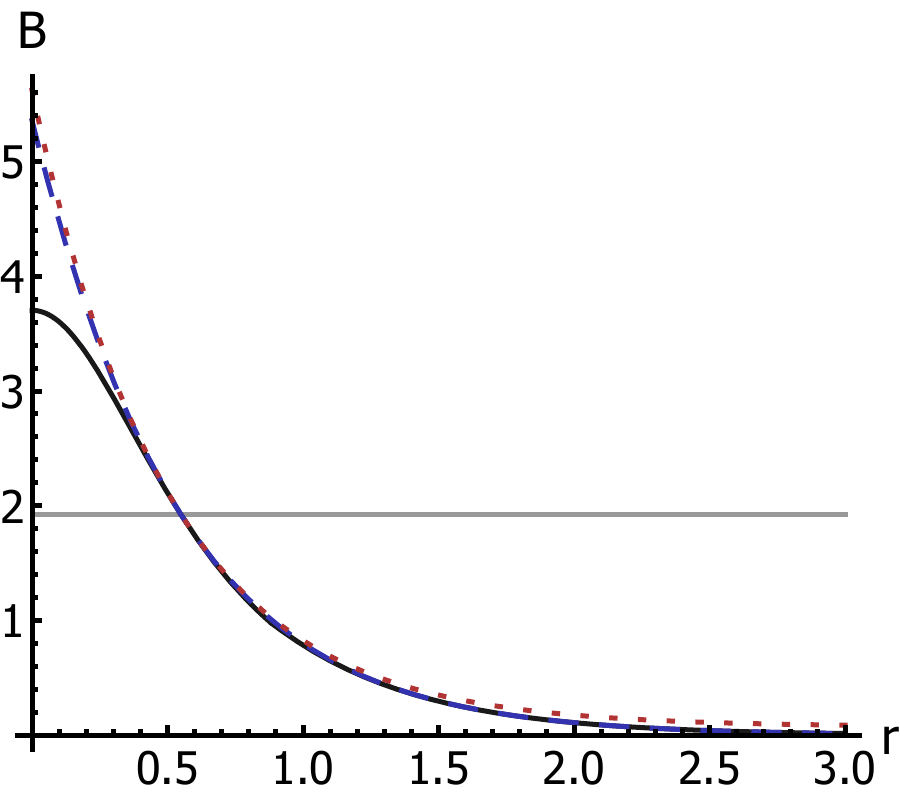}
\caption{(color online)
Dashed-blue line: average sum of the variances for two-mode squeezed thermal probe states
with $N=1$ mean thermal photons and by adopting the optimal scaling factor $K_{\rm min}$.
Dotted-red line: average sum of the variances for two-mode squeezed thermal probe states
with $N=1$ mean thermal photons and by adopting the coherent state scaling factor $K_{\rm c}$.
Solid-black line: most-informative bound for two mode squeezed thermal state (with $N=1$).
Solid-grey line: standard quantum limit ${\sf B}_{\sf SQL}$. All the plots are functions of the 
squeezing parameter $r$ and for different values of the a-priori uncertainty $\Delta$. 
Top-left: $\Delta=1$; top-right: $\Delta=2$; bottom-left: 
$\Delta=3$; bottom-right: $\Delta=5$.
\label{f:apriori}}
\end{figure}
The different results for two-mode squeezed
thermal states are shown in Fig. \ref{f:apriori}:
one observes that by using the scaling factor $K_{\rm min}$,
the estimation strategy is nearly optimal, that is the most-informative bound
is saturated for a wide range of parameters of the probe state,
and for different values of the a-priori uncertainty $\Delta$. 
One also observes that by using the simpler
scaling factor $K_c$, one still beats the SQL limit by increasing
the squeezing parameter; in particular, for zero thermal photons ($N=0$),
the entangled assisted strategy always beat the SQL for any value of 
the squeezing parameter $r$. On the other hand this strategy 
is far to be optimal for low values of $\Delta$ and for large values 
of the squeezing parameter $r$.
\subsection{Estimation of imperfect displacement operations}
Let us consider the case where the displacement operation is imperfect. 
We thus have an additional uncertainty on the parameters 
we want to estimate. We assume that the two corresponding 
values are distributed  according to a certain probability distribution
$\mathcal{P}_{{\sf err}}(q',p')$ which has mean values $q_0$ and $p_0$.
The output state, after the displacement operation, can thus be 
written as
\be
\varrho = \int dq' \, dp' \ \mathcal{P}_{{\sf err}}(q',p') \hat{D}(q',p') \varrho \hat{D}^\dagger (q',p').
\ee
For the sake of simplicity, let us consider that the error probability is a product of 
two Gaussian independent probability distributions, {\em i.e.}
$$
\mathcal{P}_{{\sf err}}(q',p') = \mathcal{G}_{q_0, \Delta_q^2}(q') \, \mathcal{G}_{p_0,\Delta_p^2}(p').
$$
Using our entanglement-assited estimation strategy, 
we obtain the following result for the variances of the estimated paramaters:
\begin{align}
{\rm Var}(q_0) + {\rm Var}(p_0) = 2 \left( 2N+1\right)e^{-2r}  + \Delta_q^2 + \Delta_p^2.
\end{align}
It is clear that the additional uncertainties are simply added to the 
previous results, giving, as expected, a worse performance in terms of 
estimation precision.
\section{The role of entanglement}\label{s:5ent}
In our measurement scheme, we make use of entangled
Gaussian states showing EPR correlations, as probe states. 
One may then ask whether the entanglement
of these states is necessary (or even sufficient) to beat the SQL
bound obtained by means of the single-mode strategy.
For this purpose, we consider a generic two-mode Gaussian state, without local
squeezing, {\em i.e.} with $\langle (\Delta \hat{q}_i)^2\rangle = \langle (\Delta\hat{p}_i)^2\rangle$,
$i=1,2$. This is a reasonable choice because we know that local squeezing does not help
in our scheme. 
Given a generic two-mode quantum state,
the corresponding quadrature operators $\hat{q}_i$ and $\hat{p}_i$ and an
arbitrary (nonzero) real number $a$, if we define the
operators $\hat{u}$ and $\hat{v}$ as
$
\hat{u} = |a| \hat{q}_1 + \frac{1}{a} \hat{q}_2 \, ,\qquad
\hat{v} = |a| \hat{p}_1 - \frac{1}{a} \hat{p}_2\,,
$
Duan et al. \cite{Duan} proved that, the condition
\be
\langle (\Delta u)^2\rangle +\langle (\Delta v)^2 \rangle  < a^2 + \frac{1}{a^2} \label{eq:DuanEnt}
\ee
is a sufficient condition for inseparability. \\
One can easily notice that, the inseparability condition is the same as
${\rm Var}(q_0) + {\rm Var}(p_0) < {\sf B}_{\rm sql}$, 
assuming $a=1$, which gives the lowest bound in Eq. (\ref{eq:DuanEnt}).
This clearly shows that the entanglement of the probe state
is a necessary condition if we are to beat the SQL obtained using
coherent states and heterodyne measurements.
Moreover, for symmetric states, such as the two-mode
squeezed thermal state $\varrho = \hat{S}_2(r) (\nu_N \otimes \nu_N) \hat{S}_2^\dag(r)$,
it is proved that the condition (\ref{eq:DuanEnt}) with $a=1$
is a necessary and sufficient condition for inseparability \cite{Duan}. As a consequence,
for this class of states, entanglement is not only necessary but also sufficient to
beat the SQL.
It is also straightforward to find a counterexample in order to prove that in the asymmetric
case, entanglement is only necessary but not sufficient. Let us consider an
asymmetric two-mode squeezed
thermal state $\varrho=\hat{S}_2(r) \nu_{N_1}\otimes\nu_{N_2} \hat{S}_2^\dag(r)$, with
$N_1 \neq N_2$; if we set $N_1=0$ the state is always entangled for $r\neq0$, but to beat the SQL
on the estimation of displacement, one can show that $N_2$ has to be moderately low
(one can derive the threshold value as a function of squeezing parameter $r$). \\
It is worth stressing the fact that the state must be entangled before the
application of the displacement operator. If we consider the case where a two-mode squeezer
is applied after the action of the displacement operator on a thermal state $\nu_N$,
no enhancement in the precision estimation can be achieved. Here, this
squeezing operation can be thought as a part of the measurement process.
The ultimate precision in this case coincides with the results described for single-mode states and has to comply with the SQL.
This result is in fact related to the security of the CV quantum key distribution
protocol with coherent states \cite{GrangierCrypto}.

\section{Parameter estimation and uncertainty relations}\label{s:6uncrel}
 We have observed that it is possible to measure the two conjugate
parameters below the SQL. This seems to contradict with the
(generalized) Heisenberg uncertainty relations. Nevertheless, if one looks carefully
at the setup, one notices that the fundamental uncertainty  relations are
never violated: the variances corresponding to the true quantum quadrature
operators $\hat{q_i}$ and $\hat{p_i}$, on each mode involved and at every step
of the measurement setup always satisfy the uncertainty relation, as it ought to be. 
The generalized uncertainty relations derived in \cite{Arthurs} show
that an inherent and unavoidable extra noise has to be taken into account if one
wants to estimate two conjugate parameters by means of a joint
measurement. However that analysis did not take into account the possibility of having a two-mode entangled
state as the initial probe as described in the previous scheme \cite{noteGUR}.
In fact, in our
setup the pre-existent entanglement is exploited in order to perform
precise measurements on different modes, and thus on commuting observables.
Specifically, if we consider the product of the corresponding variances on the estimation
of the parameters $q_0$ and $p_0$, we are led to conclude that  the generalized
uncertainty relation seems to be violated when
\be
{\rm Var}(q_0){\rm Var}(p_0) < 1. \label{eq:GHR}
\ee
If ${\rm Var}(q_0) = {\rm Var}(p_0)$, as it is always the case by considering
$\varrho_0 = \hat{S}_2(r) \nu_N \otimes \nu_N \hat{S}_2^\dag(r)$ as a probe state and
our measurement setting, one can clearly observe that the condition (\ref{eq:GHR}) is equivalent to beating the
SQL bound. Then, as described in the previous section, entanglement is always necessary and, in the symmetric case,
also sufficient to violate the generalized uncertainty relation on the conjugate parameters by means of the proposed setup.
\section{Remarks} \label{s:remarks}
The estimation of the two conjugate parameters of a displacement
operation is important both for applications and fundamental reasons.
Displacement operations are indeed ubiquitous in most of the quantum protocols
for CV systems. On the other hand, as we remarked earlier, this estimation follows from the  uncertainty relations, and thus from the foundational
properties of quantum mechanics.
In this Letter we have presented a measurement scheme which estimates
accurately the two real parameters characterizing the unitary operation of displacement in phase space,
by using Gaussian entangled probe states and homodyne detections. 
We have further derived the ultimate quantum bounds on the
multiparameter estimation for single and two-mode input Gaussian states,
showing that our setup is optimal for a large range of parameter values
characterizing the probe states. We have discussed the role of entanglement,
showing that in our setup its presence is always necessary, and in symmetric
cases  also sufficient, to beat the standard quantum limit achievable by using
coherent input states and heterodyne detection. Finally we have analyzed in detail
the relationship between our results and the generalized Heisenberg uncertainty
relation for conjugate parameters.
\section*{Acknowledgments}
The authors acknowledge useful discussions with M. Barbieri, S. Braunstein, R. Demkowicz-Dobrzanski, M. Guta, F. Illuminati, A. Monras, S. Olivares and A. Serafini.  This work was supported by the UK EPSRC, the IT-MIUR (FIRB RBFR10YQ3H), and the NPRP 4-554-1-084 from Qatar National Research Fund.

\end{document}